\documentclass[aps,prb,reprint,superscriptaddress]{revtex4-2}
\usepackage{hyperref}
\usepackage{graphicx}
\usepackage{color}
\usepackage{amsfonts}
\usepackage{amsmath}
\usepackage{physics}

\usepackage[utf8]{inputenc}
\usepackage[T1]{fontenc}
\usepackage{mathtools}
\usepackage{bbm}
\usepackage[dvipsnames]{xcolor}
\hypersetup{colorlinks=true, urlcolor=blue, citecolor = blue}
\usepackage[english]{babel}

\begin{document}

\title{Spin and orbital mixing of edge states\\ in a quantum Hall system proximitized by a superconductor}

\author{S. Maji}
\email{maji@agh.edu.pl}
\affiliation{AGH University of Krakow, Academic Centre for Materials and Nanotechnology, al. A. Mickiewicza 30, 30-059 Krakow, Poland}

\author{M. P. Nowak}        
\email{mpnowak@agh.edu.pl}
\affiliation{AGH University of Krakow, Academic Centre for Materials and Nanotechnology, al. A. Mickiewicza 30, 30-059 Krakow, Poland}

\date{\today}

\begin{abstract}
We investigate the formation and transport properties of chiral Andreev edge states in a two-dimensional quantum Hall system proximitized by a superconductor. By numerically modeling the system using the Bogoliubov-de Gennes equations, we analyze the non-local conductance and transmission probabilities of multimode and spinful systems. We demonstrate that the Andreev reflection process induces a mixing of the quantum Hall edge modes at higher filling factors, a phenomenon strictly prohibited in clean, purely electronic systems. When incorporating the Zeeman interaction, we show that the Andreev edge states split into uncoupled spin species, maintaining spin orthogonality that prevents mixing between opposite spin sectors. Furthermore, we explore the impact of Rashba spin-orbit coupling. While the spin-orbit interaction alone causes slight spin depolarization, its combination with an in-plane magnetic field drives complex spin mixing among all chiral Andreev bands, fundamentally altering the conductance oscillations. Finally, we reveal that the electron transmission probabilities exhibit robust degeneracies, which emerge as a direct consequence of the unitarity constraints and the particle-hole symmetry of the system's scattering matrix in a magnetic field and the presence of spin-orbit interaction.
\end{abstract}

\maketitle
\section{Introduction}
The integration of semiconductor and superconductor hybrid nanostructures within the quantum Hall (QH) regime has spurred significant interest in recent years as a promising platform for designing novel topological phases \cite{Mong, Amet, Lee2017, Ronen, Anjana, Clarke2014, Hoppe}. Due to the proximity effect driven by Andreev reflection at the normal-superconductor interface \cite{Klapwijk2004}, an incident electron with energy within the superconducting gap is reflected as a hole, forming a Cooper pair in the superconductor \cite{osti_4071988}. Along the QH-superconductor interface, this process leads to the formation of Chiral Andreev Edge States (CAES) as a result of skipping orbit trajectories, where an electron continuously converts into a hole and back into an electron upon successive Andreev reflections. Under specific conditions, CAES are proposed to manifest as self-conjugate Majorana fermions \cite{Chamon,Gamayun,Gaurav,Rakesh,v68s-tgxs}. Unlike conventional fermions, these exotic quasiparticles follow non-abelian anyonic exchange statistics, making them a crucial ingredient for designing topological qubits---the foundational building blocks for fault-tolerant quantum computation \cite{Alicea, KITAEV20032, Nayak, Maji}. This search has motivated intensive efforts in studies of semiconductor-superconductor structures, particularly in the QH regime, as a natural way of realizing quasi-one-dimensional edge modes.

However, the successful coexistence of the QH effect in a two-dimensional electron gas (2DEG) with superconductivity presents a formidable experimental challenge, as the high magnetic fields required to achieve Landau level quantization typically suppress superconducting correlations. Initially, graphene emerged as a possible route to overcome these challenges, as it manifests the quantum Hall effect at relatively low magnetic fields (below 1 T) and is compatible with superconducting materials \cite{Lee2017,Park2017,Ronen, Borzenets}. Despite this advantage, the complex Landau level structure and the small Land\'e $g$-factor of graphene make the detailed study of spin dynamics in these systems challenging. Fortunately, recent experiments have provided strong evidence for the survival of the superconducting proximity effect at highly transparent QH-superconductor interfaces within 2DEGs under high magnetic fields \cite{TAKAYANAGI1998462, Eroms, Wan2015, Calado2015, BenShalom2016} and signatures of Andreev reflections \cite{Cuozzo, Akiho, Kanter}.

Motivated by this development, in this work, we are particularly interested in the interplay between the QH effect and induced superconductivity in a 2DEG-superconductor device, with an emphasis on the effects of spin interactions provided by the semiconductor, such as the Zeeman splitting and Rashba spin-orbit interaction (SOI). We consider a widely utilized multi-terminal Hall-bar geometry \cite{Klitzing} where charge carriers form conducting chiral edge channels at the boundary of an insulating bulk, with a superconducting contact attached to one of the terminals.  

While previous literature has explored the formation of CAES in the spin-degenerate regime \cite{Zhao2020, 10.21468/SciPostPhysCore.5.3.045}, proposals of chiral Majorana bound states \cite{Qi, Wang}, and even signatures of their absence \cite{Uday,Morteza}, here we extensively study these hybridized electron-hole states by solving the Bogoliubov-de Gennes (BdG) equations specifically for multimode and spinful systems \cite{Takagaki, Eroms, Khaymovich}. We explore the underlying mechanisms of spin and orbital mixing of these edge states, shedding light on the complex transmission symmetries and the profound effects of the Zeeman interaction and spin-orbit coupling on chiral transport at the proximitized boundary. We demonstrate that the mixing between CAES states is induced by Andreev reflection in the presence of more than one orbital QH edge mode, as well as by spin mixing in the presence of SOI. Importantly, we show that the electron transmission coefficients obey particular degeneracies, which directly result from the symmetry of the system's scattering matrix, ensured by unitarity and particle hole symmetry.

The paper is organized as follows. Section II provides the theoretical background to the conducted study. The results for spin-degenerate systems are provided in Sect. III A. Sect. III B considers the effects of Zeeman interaction, while Sect. III C discusses spin mixing introduced by SOI and transmission matrix symmetries. Section IV concludes the paper.

\section{Theory}
\begin{figure}
    \centering
    \includegraphics[width=0.98\linewidth]{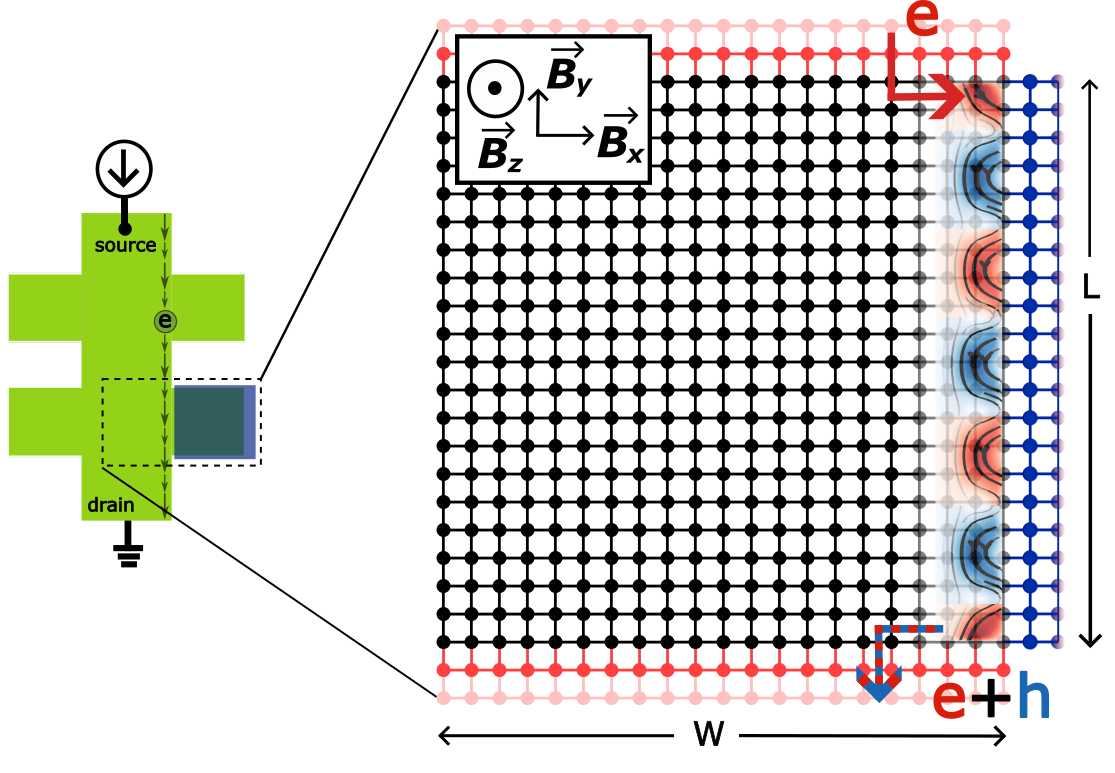}
    \caption{The scheme of the considered system. A semiconducting Hall bar (green) in the presence of perpendicular  magnetic field hosts edge states. Blue region denotes a superconducting contact. Zoom-in shows the system considered in the numerical calculation. On top and bottom we include two normal semi-infinite leads (red), while on the right edge we connect semi-infinite superconducting contact (blue). The electrons propagating on the QH edge mode forms CAES at the normal-superconductor interface. We plot resulting probability current at the edge, where colors correspond to its electron (red) and hole (blue) character. For visual clarity, the mesh shown in the figure is coarser than the one used for numerical calculation.}
    \label{fig:system_final}
\end{figure}
 
To study the formation of CAES on the normal-superconductor edge, we focus on part of the entire Hall bar, as depicted in Fig. \ref{fig:system_final}. The system is described by the Hamiltonian written in the basis $(\psi_{e\uparrow},  \psi_{h\downarrow}, \psi_{e\downarrow}, -\psi_{h\uparrow})^T$, is,
\begin{equation}
\begin{split}
\label{eqn:Hamiltonian_Zeeman_SO}
H = &\left(\frac{\hbar^2 \mathbf{k}^2}{2m^*} - \mu\right)\sigma_0\otimes\tau_z + \Delta(r)\sigma_0\otimes\tau_x +\\&+ \alpha(r)(\sigma_x k_y - \sigma_y k_x)\otimes\tau_z + \frac{1}{2}g(r)\mu_B\mathbf{B}\mathbf{\sigma}\otimes\tau_0,
\end{split}
\end{equation}
where $\mathbf{k} = -i\nabla_r$, $\sigma_j$ and $\tau_i$ with $i = x, y, z$ are the
Pauli matrices that act on the spin and electron-hole degree
of freedom, respectively.  $\frac{1}{2}g\mu_B\mathbf{B}\mathbf{\sigma}\otimes\tau_0$ is the Zeeman interaction term, with $g$ being the Lande $g$-factor and the magnetic field of arbitrary orientation $\mathbf{B} = [B_x, B_y, B_z]$. $\mu$ is the chemical potential, and $\alpha$ is the Rashba SOI amplitude. The Zeeman interaction and the spin-orbit coupling are included in the normal part of the system, along with the perpendicular magnetic field for the QH effect. The perpendicular magnetic field is introduced by the substitution $\hbar k \rightarrow \hbar k - qA(r)$, where $\vec{A}$ is the magnetic vector potential chosen in the Landau gauge $\textbf{A} = (0, B_zx, 0)$. In the tight binding description, we discretize the Hamiltonian on a square lattice with a lattice constant $a$ (see the system plot depicted in the right part of Fig. \ref{fig:system_final}), and the orbital effects are implemented by adding a gauge-dependent phase factor in the hopping term $t_{ij}\rightarrow t_{ij}\exp[\frac{\mp ie}{\hbar}\int_{r_j}^{r_i} \vec{A} \cdot d\vec{l}]$, with the sign being opposite in the electron/hole sector. The pairing potential present in the superconductor is included in the Hamiltonian by the term $\Delta$.  

We adopt the system shape: length $L = 1000$ nm, width $W = 1000$ nm, the normal-superconductor interface located at $x= 0$, and assume InSb material parameters $m^*=0.014m$, $g =-50$, $\alpha = 50$ meVnm, and the induced superconducting gap of $\Delta = 2$ meV, corresponding to superconductors such as Nb or NbTiN \cite{Salimian, Zhi, Jinhua, Satchell}. However, the conclusions and phenomena described in this work are not specific to the choice of this particular set of parameters, and choosing different materials would rather scale the significance of particular mixing processes, as the strengths of the Zeeman interaction, SOI would be different.

To analyze the transport properties of the system, we adopt the Landauer-Buttiker approach. The conductance of the system is obtained via the formula:
\begin{equation}
G_{ij}(E) = \frac{\partial I_i}{\partial V_j} = \frac{e^2}{h}(\delta_{ij}N_i^e(E)-T^{ee}_{ij}(E)+T^{he}_{ij}(E)).
\label{conductanceformula}
\end{equation}
$I_i$ is the current entering the scattering region from the terminal $i$, and $V_j$ is the voltage applied to the $j$'th lead. $N_i$ is the number of electronic modes in the $i$'th lead; $T^{ee}_{ij}(E)$ is the electron-to-electron transmission coefficient for electrons injected from the lead $j$ and captured at the terminal $i$; and $T^{he}_{ij}(E)$ is the corresponding electron-to-hole transmission coefficient. Here, we focus on zero-temperature non-local conductance by placing $i \ne j$ and considering injecting the electrons from the top lead such that they create edge modes on the right side of the system in a positive perpendicular magnetic field. In this work, $j$ corresponds to the top lead from which the electrons are injected, and $i$ is the bottom lead where the outgoing quasi-particles are captured; hence the corresponding conductance becomes $G=G_{\mathrm{bottom}, \mathrm{top}}$.

The coefficients $T^{ee}_{ij}(E)$ and $T^{he}_{ij}(E)$ are obtained from the scattering matrix of the system calculated at the energy $E$ corresponding to the voltage bias on the $j$'th lead,
\begin{equation}
    T^{\alpha,\beta}_{kl}(E) = \mathrm{Tr}\left( [S^{\alpha,\beta}_{kl}(E)]^\dagger S^{\alpha,\beta}_{kl}(E)\right).
\end{equation}
$S^{\alpha,\beta}_{kl}(E)$ is the block of the scattering matrix corresponding to the particles of type $\beta$ injected from the $l$'th lead and scattered back as the particle type $\alpha$ into the $k$'th lead. We particularly focus on a situation with negligible voltage bias and set $E=0$. 

The scattering matrix of the system is obtained by discretizing the Hamiltonian Eq. (\ref{eqn:Hamiltonian_Zeeman_SO}) on a mesh with lattice spacing $a = 5$ nm and solving the transport problem using the Kwant package \cite{Christoph}. Conductance maps are calculated with the help of the Adaptive Python package \cite{nijholt_2023_10215599}. The code used to obtain the results presented in this paper is available in an online repository \cite{maji_2026_20271456}.

\section{Results}
In this section, we study the orbital effects and spin effects, respectively. We first consider the spin degenerate case with $\alpha = 0$, $g=0$ in subsection III A. Only the perpendicular magnetic field is included in order to incorporate the QH effect. In subsection III B, we include the spin effect of magnetic field through the Zeeman interaction with $g= - 50$, $\alpha = 0$. Then, in subsection III C, finally the Rashba spin-orbit coupling is included with $\alpha = 50$ meV nm. 
\subsection{Spin-degenerate case}
\begin{figure}
    \centering
    \includegraphics[width=0.75\linewidth]{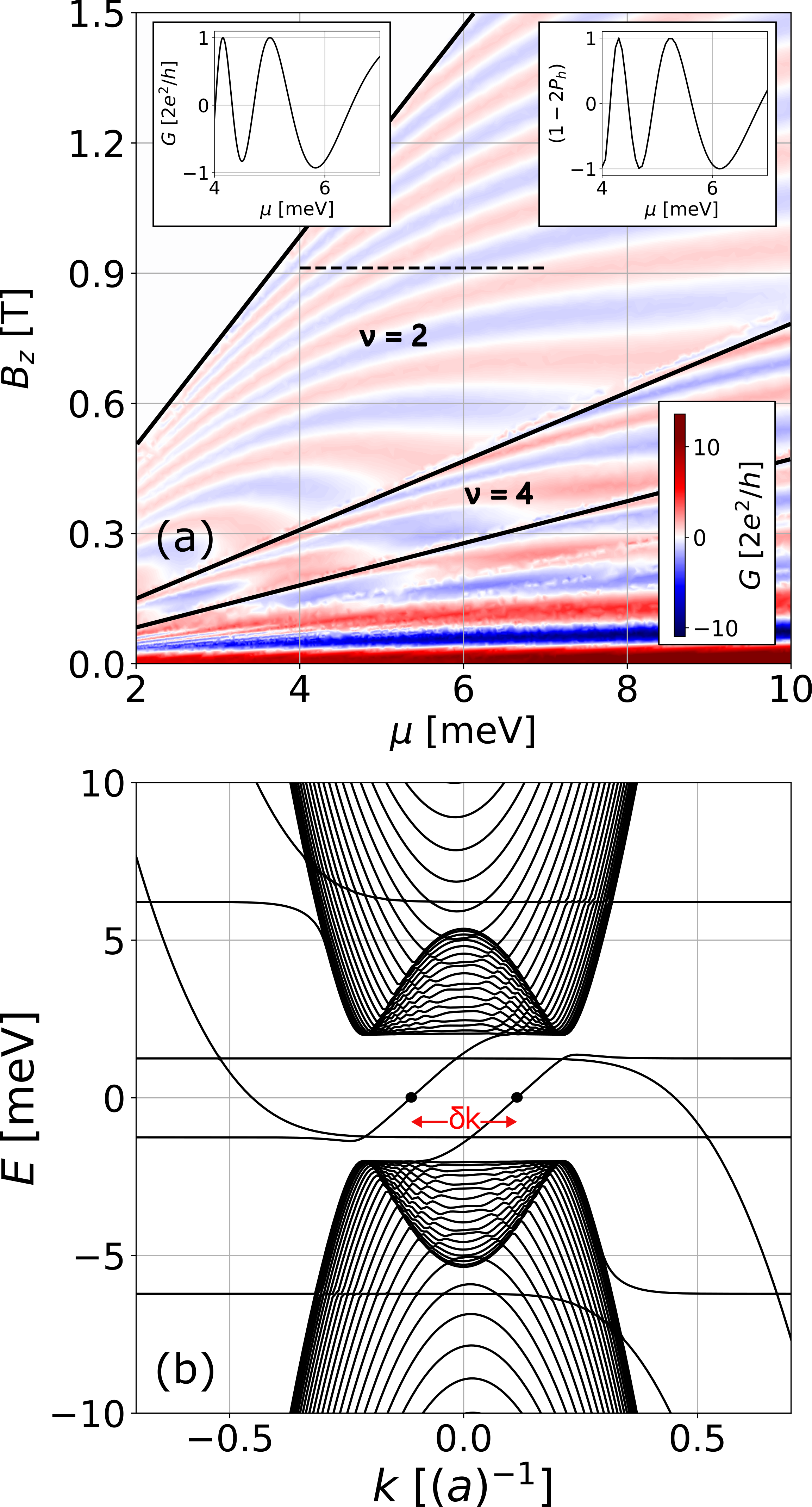}
    \caption{(a) Conductance of normal-superconductor system in the QH regime versus the perpendicular magnetic field magnitude and chemical potential. Black lines separate the regions with different filling factor $\nu$. The left inset of the plot shows a conductance cross-section at $B_z = 0.912$ T, $\alpha = 0$, $g=0$, obtained for the $\mu$ values denoted in the conductance map, right inset shows the probability calculated from the analytical formula Eq. \ref{eqn:probability}. (b) Band-structure of proximitized region with the dots denoting the CAES bands crossings zero energy obtained for $\mu = 5$ meV and $B = 0.912$ T, $\alpha = 0$, $g=0$. Wave vector difference of the CAES pair is denoted by $\delta k$.}
    \label{fig:G_spinless}
\end{figure}

To describe the effects of mode mixing, let us first assume that there are no spin-interactions and treat the system as spin degenerate. Here, the effects of the magnetic field are included solely through the Peierls phase. Figure \ref{fig:G_spinless}(a) shows the conductance of the proximitized Hall-bar region. The black lines separate regions with different filling factor values $\nu$, which, due to spin degeneracy, are always even and increase by a factor of 2. A clear signature of oscillation is observed in each distinct regime of $\nu$. To further inspect the oscillations, we take a cross-section from the conductance plot at a constant magnetic field $B_z = 0.912$ T (marked with a black dashed line in Fig. \ref{fig:G_spinless}(a)) and show it in the left inset of Fig. \ref{fig:G_spinless}(a).

The oscillations of the conductance are the result of the interference of two CAES states. The presence of them at the normal-superconductor interface can be visualized by plotting the dispersion relation of the proximitized QH system with assumed translational invariance along the interface (for calculations of all dispersion relations shown in this paper we replace the superconducting lead with a finite superconducting region), which makes $k_y$ a good quantum number. We consider a chemical potential $\mu = 5$ meV to isolate the case $\nu = 2$ and show the resulting dispersion relation in Fig. \ref{fig:G_spinless}(b). In the dispersion, we observe a set of parabolic bands corresponding to quasiparticle states in the superconductor, separated by the induced gap $2\Delta$. Inside the gap, we identify electron-hole hybridized CAES emerging from the interplay between the chiral QH edge states and the Andreev reflection process at the superconductor interface.

For a spinless edge state, the superconducting proximity effect hybridizes the electron-hole edge channel and can be described by the BdG Hamiltonian in the basis of $\{ \ket{e}, \ket{h} \}$. Diagonalizing this Hamiltonian at zero energy yields two distinct orthogonal eigenstates $\ket{\psi_1}$ and $\ket{\psi_2}$. Due to the intrinsic particle-hole symmetry, the two eigenstates at zero energy can be represented by \cite{Zhao2020} 
\begin{equation}
\begin{split}
\label{eqn:wave_functions}
\ket{\psi_1}&=\alpha\ket{e}+\beta\ket{h}\\
\ket{\psi_2}&=\beta^*\ket{e}-\alpha^*\ket{h},
\end{split}
\end{equation}
where, $|\alpha|^2+|\beta|^2 = 1$. The probability of an electron turning into a hole after traveling through the proximitized edge is $P_h=4|\alpha|^2|\beta|^2\sin^2({\delta k L/2})$, where $\delta k = k_1-k_2$ at zero energy. $k_1$, $k_2$ are the wave-vectors of the two modes, and $L$ is the propagation length. Therefore, the electron-hole probability difference that comes from the particle traveling within the two CAES after a length $L$ is
\begin{equation}
\label{eqn:probability}
P_e-P_h=(1-2P_h)=(1-8|\alpha|^2|\beta|^2\sin^2({\delta k L/2})).
\end{equation}
This probability directly translates to non-local conductance, which is positive when the injected electron escapes the system as an electron and negative when the hole escapes the system instead. The oscillation shown in the left inset of  Fig. \ref{fig:G_spinless}(a) can be explained using the analytical formula of Eq. \ref{eqn:probability}. We extract the $\delta k$ values varying $\mu$ from the dispersion relation of the system at zero energy. In the right inset of Fig. \ref{fig:G_spinless}(a), we see that the accumulated phase difference $\delta k L$ along the path of CAES propagation causes the interference, and hence, the electron-hole probability difference oscillations are compatible with those obtained numerically. 
The small relative shift can be attributed to a slight variation in the actual travel distance along the length of the interface $L$ when the electrons are injected from and captured in the normal leads perpendicular to the interface.

\subsubsection{Mode mixing in $\nu > 2$ case}
\begin{figure}
    \centering
    \includegraphics[width=0.9\linewidth]{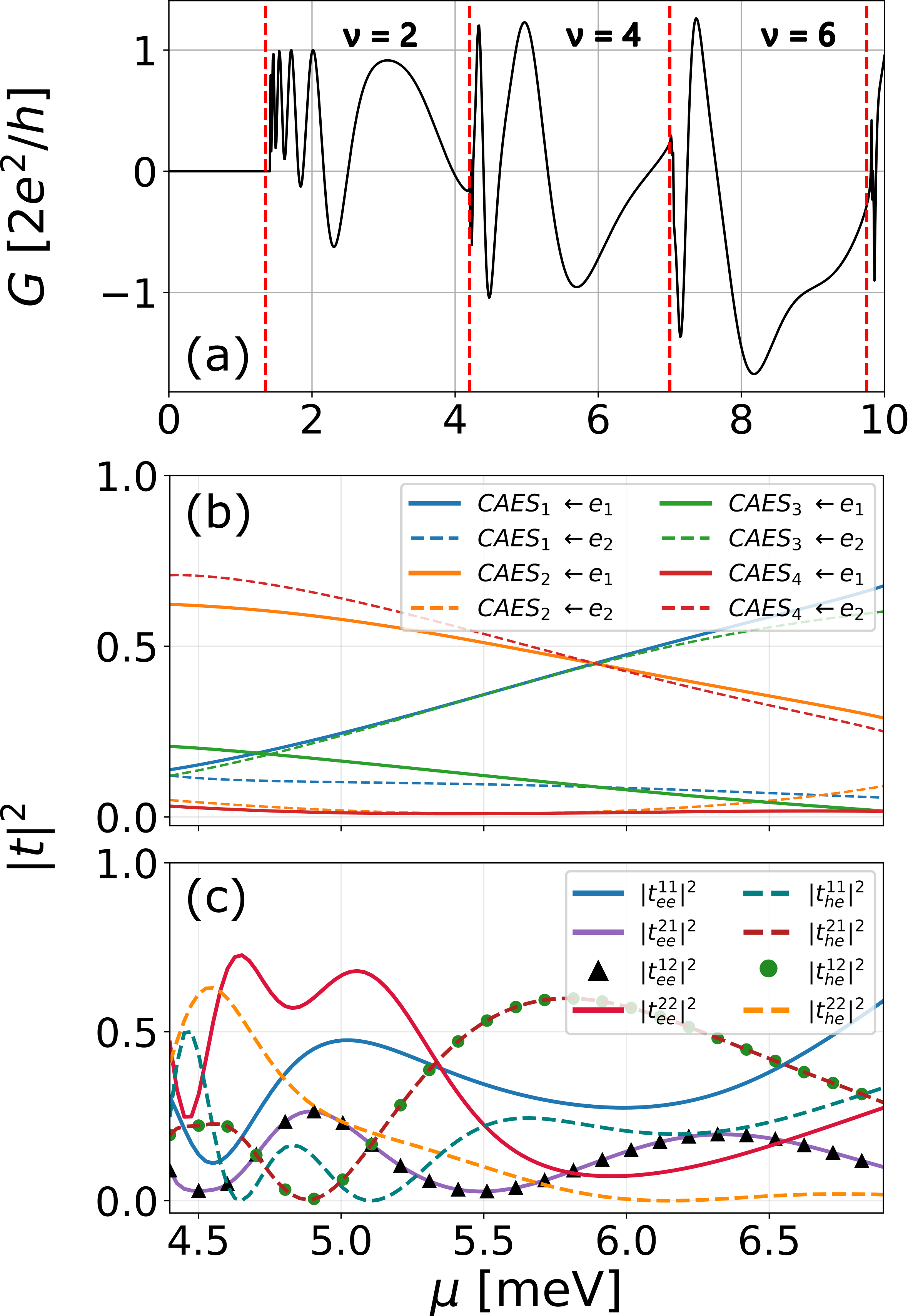}
    \caption{(a) Conductance cross-section versus chemical potential $\mu$ marked with different $\nu$ regimes. (b) Occupation probability of different CAES states for different initial electron edge modes in $\nu = 4$ regime. (c) Components of electron and hole transmission probability in $\nu = 4$ regime. The perpendicular magnetic field is $B_z = 0.343$ T, $\alpha = 0$, $g=0$.}
    \label{fig:QH_SC_nu_2}
\end{figure}

Let us move to the regime of higher filling factors. We set the magnetic field $B_z = 0.343$ T and now varying $\mu$ drives us through different $\nu$ values. In this limit, we see more complex conductance oscillations (see Fig. \ref{fig:QH_SC_nu_2}(a)), which cannot be explained simply by the analytical formula Eq. \ref{eqn:probability}. 

To understand this behavior, we focus on the $\nu = 4$ case. We calculate the probabilities of an electron originating from two available (each spin degenerate) QH channels in the normal lead being injected into four possible CAES. We define the
transmission probability
\begin{equation}
\label{eqn:probability_CAES}
    P_{i,j} = |\mathcal{T}_{i,j}|^2,
\end{equation}
the probability that a particle injected in incoming mode $j$ is transmitted
into outgoing CAES mode $i$. $\mathcal{T}_{ij}$ is the transmission matrix element of a system in which the bottom lead is replaced by a lead with a normal and SC segment.
The results are shown in Fig. \ref{fig:QH_SC_nu_2}(b). For clarity, we label the transmission probability $P_{ij}$ as $CAES_i \leftarrow e_j$ in the plot. The solid lines correspond to probabilities for an electron entering the system in the first edge mode. We see that as the chemical potential is varied, the electron that is injected from this mode couples to three CAES with quite significant probabilities. The same happens for the electron that enters the system in the second edge channel (dashed curves). This is a clear signature of Andreev reflection mixing of the QH edge modes, and therefore we cannot treat the CAES separately as coming from different edge modes, prohibiting the convolution of the total electron transport probability as resulting from two independent processes (from each edge mode) using a simple interference-based explanation of the oscillations in conductance, as done for $\nu = 2$.

The mixing of the edge mode can be further analyzed by tracing the probabilities of conversion of an incoming electron into an outgoing electron and hole on the other edge of the system, $|t_{ee}|^2$ and $|t_{he}|^2$, respectively. For more than a single edge channel in the normal lead, they are defined as $|t_{ee}^{ij}|^2$, $|t_{eh}^{ij}|^2$, where $i,j \in \{1,2\}$ correspond to the output and input quantum edge mode channels, respectively.  We plot the transmission probabilities in Fig. \ref{fig:QH_SC_nu_2}(c).

For a clean and purely electron QH system, the edge modes do not mix. Therefore, the electron entering the first edge channel must escape the system in the same mode. Here, the situation is different despite the system being perfectly transparent. The electron entering the system in the first edge mode has a non-zero probability of escaping the system in the second mode (see the violet curve for $|t_{ee}^{21}|^2$ in Fig. \ref{fig:QH_SC_nu_2}(c)). The electron injected in the second mode is also partially escaping the system in the first mode (see the black triangles for $|t_{ee}^{12}|^2$). Interestingly, we find $|t_{ee}^{21}|^2$ to be exactly equal to $|t_{ee}^{12}|^2$. This initially puzzling observation can become understandable when we note that the transport from the top lead from electron mode 1 to electron mode 2 in the bottom lead is exactly the same process as the transport from the bottom lead from mode 2 to the top lead mode 1. Those processes are identical, and their sameness is guaranteed by the time reversal symmetry of the {\it whole} system---the reverse of the propagation direction and the orientation of the magnetic field. The combined action of the time reversal operator, along with the reversal of the magnetic field orientation guarantees $TH_{B \rightarrow -B} T^{-1} = \frac{\hbar^2}{2m^*}(-k+e(-A))^2 = H$, where $T = \mathcal{K}$ is a complex conjugation time reversal operator. This means that the Hamiltonian is invariant under the application of the time reversal operation when the magnetic field is also reversed. Symmetry is also observed between the $|t_{eh}|^2$ amplitudes (see the dashed red curve and the trace plotted with the green dots). In this case, however, the electrons and holes in the same band have the same Fermi velocities as provided by particle-hole symmetry, only at zero energy. Therefore, this symmetry (changing the injected particle from electron to hole and vise versa) holds only for zero bias, as we have checked. We have also verified that for a non-ideal normal-superconductor interface, the conductance oscillations are quantitatively modified, while the underlying CAES interference mechanism discussed here remains unchanged.

\subsection{Spinful system - Zeeman interaction}
\begin{figure}
    \centering
    \includegraphics[width=0.8\linewidth]{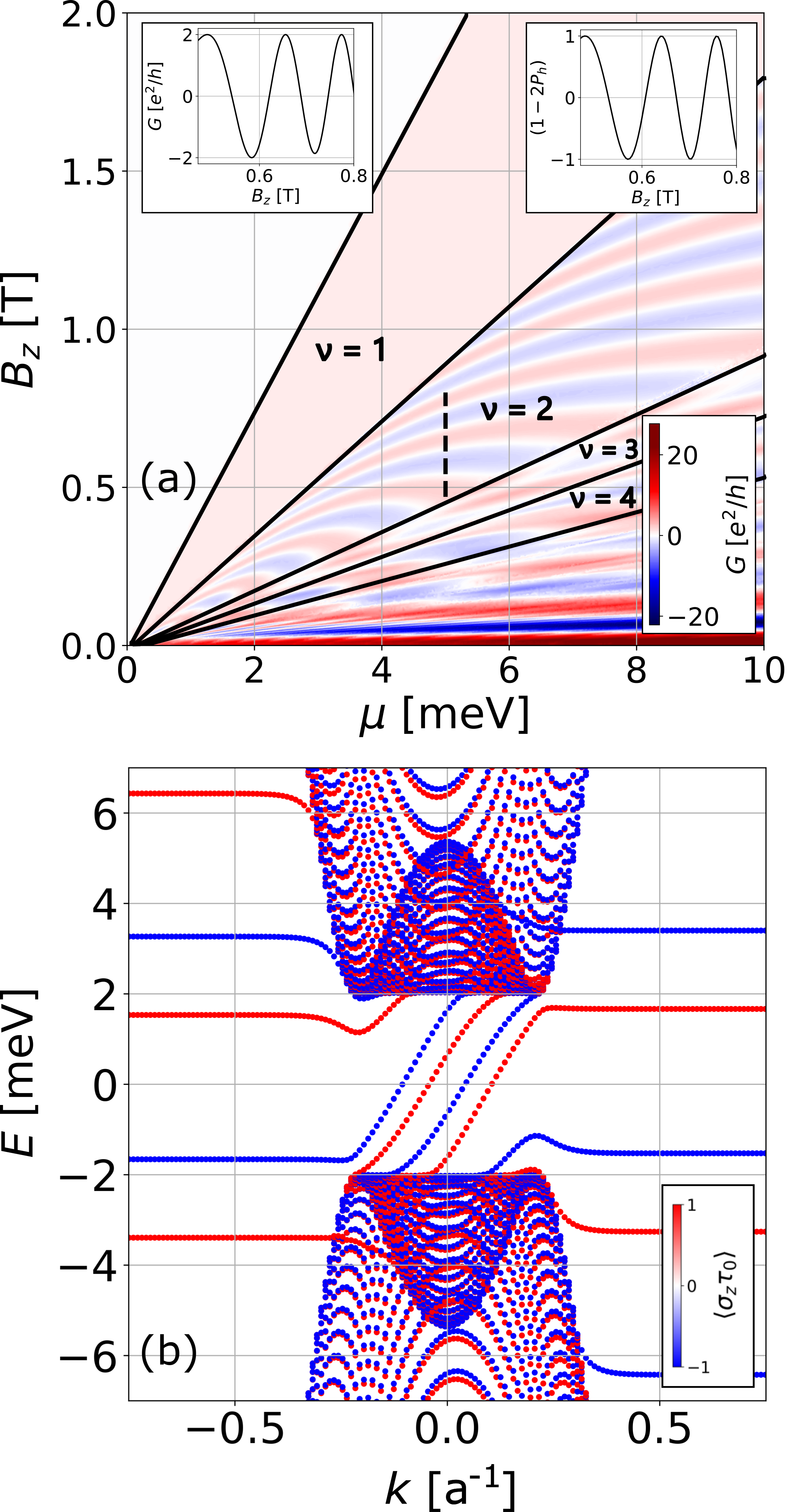}
    \caption{(a) Conductance versus perpendicular magnetic field and chemical potential with the Zeeman interaction included. The left insets show conductance cross-section versus $B_z$ at $\mu=5$~meV (left) and the probability calculated from the analytical formula Eq.~\ref{eqn:probability} (right) with, $g=-50$, $\alpha = 0$. (b) The band structure of the normal-superconducting system colored with the spin-polarization of the bands for $\mu = 5$ meV, $B_z = 0.6$ T with, $g=-50$, $\alpha = 0$.}
    \label{fig:G_Zeeman_z}
\end{figure}

For materials such as InAs and InSb, which have a considerable $g$-factor, the perpendicular magnetic field not only introduces the orbital effects that lead to the creation of edge modes but also introduces spin splitting through the Zeeman effect. Fig. \ref{fig:G_Zeeman_z}(a) shows the conductance maps obtained with the inclusion of the Zeeman spin splitting. We now observe that for a given chemical potential $\mu$, the system passes through consecutive values of the filling factor $\nu$, including both odd and even values. This is a clear signature of the Zeeman interaction splitting the spin bands. Importantly, this lack of degeneracy persists at even values of $\nu$. This can be inspected in the band structure presented in Fig. \ref{fig:G_Zeeman_z}(b). There we observe four CAES crossing the zero energy. In the plot, we color the bands by their spin expectation value $\langle \sigma_z \tau_0 \rangle$. We clearly see that there are two sets of CAES bands in the dispersion. One is created by an electron-hole pair with electron spin polarized up and hole spin polarized down (red), and a pair in which the spin is oriented oppositely (blue). This is consistent with the s-wave electron pairing assumed here. Now, if we look at the conductance map, we observe pronounced oscillations. The inset on the left of Fig. \ref{fig:G_Zeeman_z}(a) shows the cross-section of the conductance map obtained for $\mu = 5$ meV and $\nu = 2$. By extracting the difference in the wave-vector $\delta k$ at zero-energy from one of the equivalent pairs [either bands with positive (red) or negative (blue) spin polarization in Fig. \ref{fig:G_Zeeman_z}(b)] and varying the magnetic field, we calculate the probability using the analytical formula, Eq. \ref{eqn:probability}. The analytically obtained probabilities are shown in the right inset of Fig. \ref{fig:G_Zeeman_z}(a) and resemble those obtained from the numerical calculation. This means that the analytical approach, which assumes only two CAES modes, still holds and that the two pairs of CAES do not mix. This is a clear result of the spin orthogonality of CAES modes, which prohibits mixing between the opposite spin sectors of the original BdG Hamiltonian. 

Furthermore, at large magnetic fields, where the Zeeman interaction is pronounced, one observes a $\nu = 1$ region with constant conductance. This is a clear signature of the splitting of the spin bands, so below the Fermi energy, there is only one spin species available, prohibiting Andreev reflection and hence CAES creation. In this case, the conductance takes the value of one conductance quantum $e^2/h$ as a single spin polarized mode is transmitted through the system.

\subsubsection{In-plane field}
\begin{figure}
    \centering
    \includegraphics[width=0.7\linewidth]{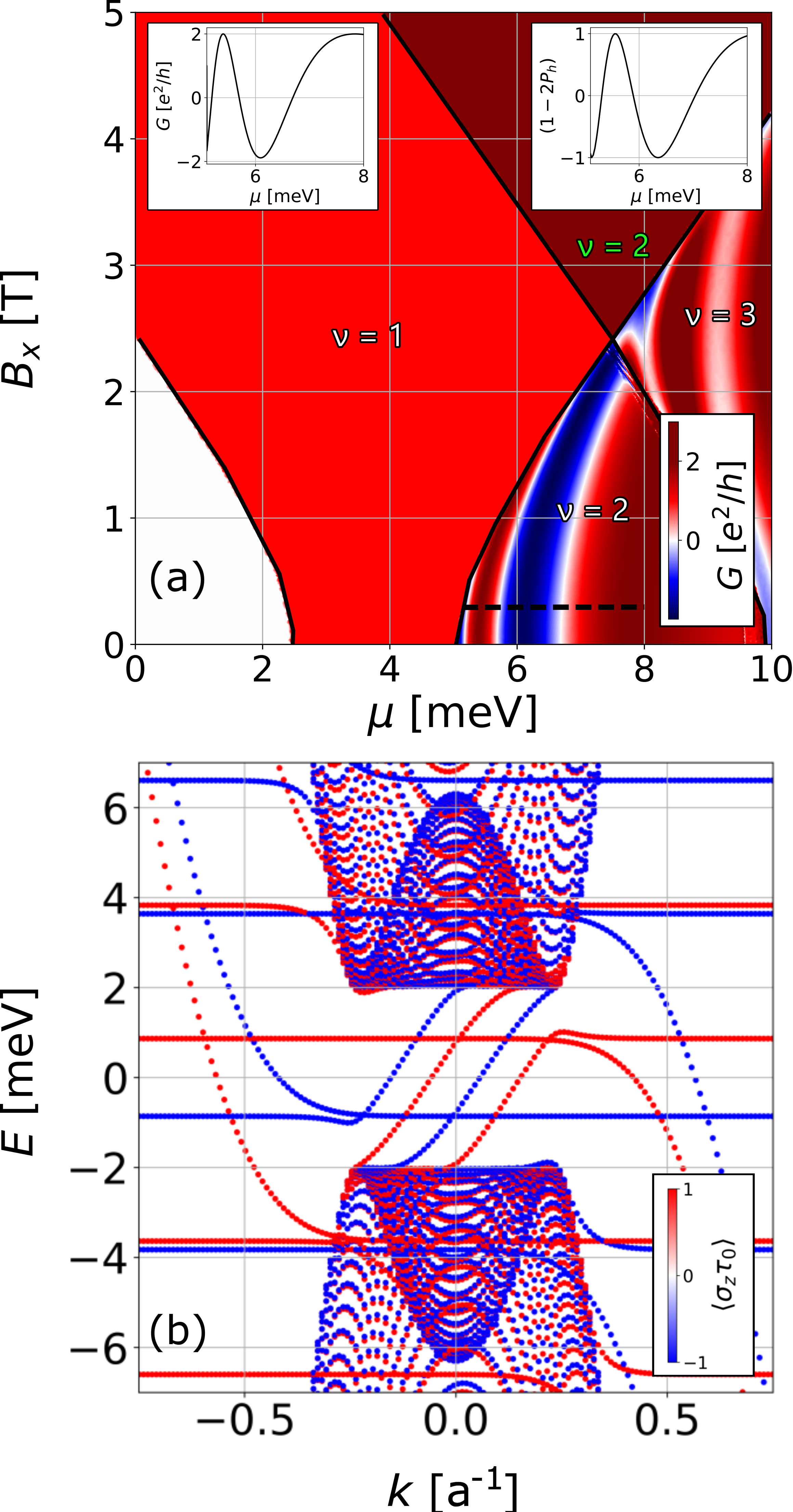}
    \caption{(a) Conductance of a QH-superconductor system versus chemical potential and the magnitude of in-plane magnetic field oriented in $x$-direction with, $g=-50$, $\alpha = 0$. The left inset of the map shows conductance cross-section versus $\mu$ at $B_x=0.3$ T, right one shows the probability calculated from the analytical formula Eq.~\ref{eqn:probability}. (b) The band structure of the system plotted with spin-polarization for $\mu = 6$ meV, $B_z = 0.912$ T, $B_x = 0.3$ T, $g=-50$, $\alpha = 0$.}
    \label{fig:G_Zeeman_x}
\end{figure}

Spin splitting in the system can also be intentionally induced by an external in-plane magnetic field. Figure~\ref{fig:G_Zeeman_x}(a) shows the conductance map obtained for a fixed perpendicular magnetic field $B_z = 0.912$ T, but with a varied magnitude of the in-plane field oriented in the $x$-direction. Since in this case the Zeeman splitting controlled by $B_x$ is decoupled from the modification of the edge band structure through the orbital effects via the perpendicular magnetic field, we can directly observe how the spin splitting modifies the transport through CAES.  In the map, we can observe distinctly different regimes: those with a constant conductance and those where the conductance oscillates (denoted by the appropriate filling factors $\nu$). When $B_x$ is small or the chemical potential is large, there are at least two spin opposite modes below the Fermi energy, and we observe conductance oscillations. Since the presence of the magnetic field in the $z$ and $x$ directions results in a well defined spin, we again can still find CAES bands that are grouped into pairs of opposite spins. We plot the band structure for $\nu = 2$ in Fig. \ref{fig:G_Zeeman_x}(b). Following the same procedure as previously, we extract the wave-vector difference for each spin-polarized pair, and in the insets to Fig.~\ref{fig:G_Zeeman_x}(a), we show the comparisons of the numerically obtained conductance and electron-hole conversion probability. As they match, we again find that sole spin splitting results in the creation of two uncoupled species of CAES. 

An interesting case occurs for $\nu = 3$. Here we have two spin-opposite bands of the same edge mode that result in the creation of CAES and one-spin polarization without the partner. The mode mixing provided by Andreev reflection discussed in the previous section results in the creation of 6 CAES bands and pronounced conductance oscillations. A different situation occurs for a small chemical potential ($\nu = 1$) or a larger magnitude of $B_x$ and larger $\mu$ ($\nu = 2$ marked in green). There, the conductance is constant and equal to the number of electron edge modes. Here, the strong Zeeman splitting of the bands leaves only electron bands with the same spin polarization below the Fermi energy, prohibiting Andreev reflection and hence the creation of CAES. 

Exactly the same results are obtained if the Zeeman interaction is applied in $y$ or in any other in-plane direction instead of $x$, as it only varies the direction of spin polarization of the modes without affecting CAES creation.

\subsection{Spin mixing in the presence of SOI}
\begin{figure}
    \centering
    \includegraphics[width=0.7\linewidth]{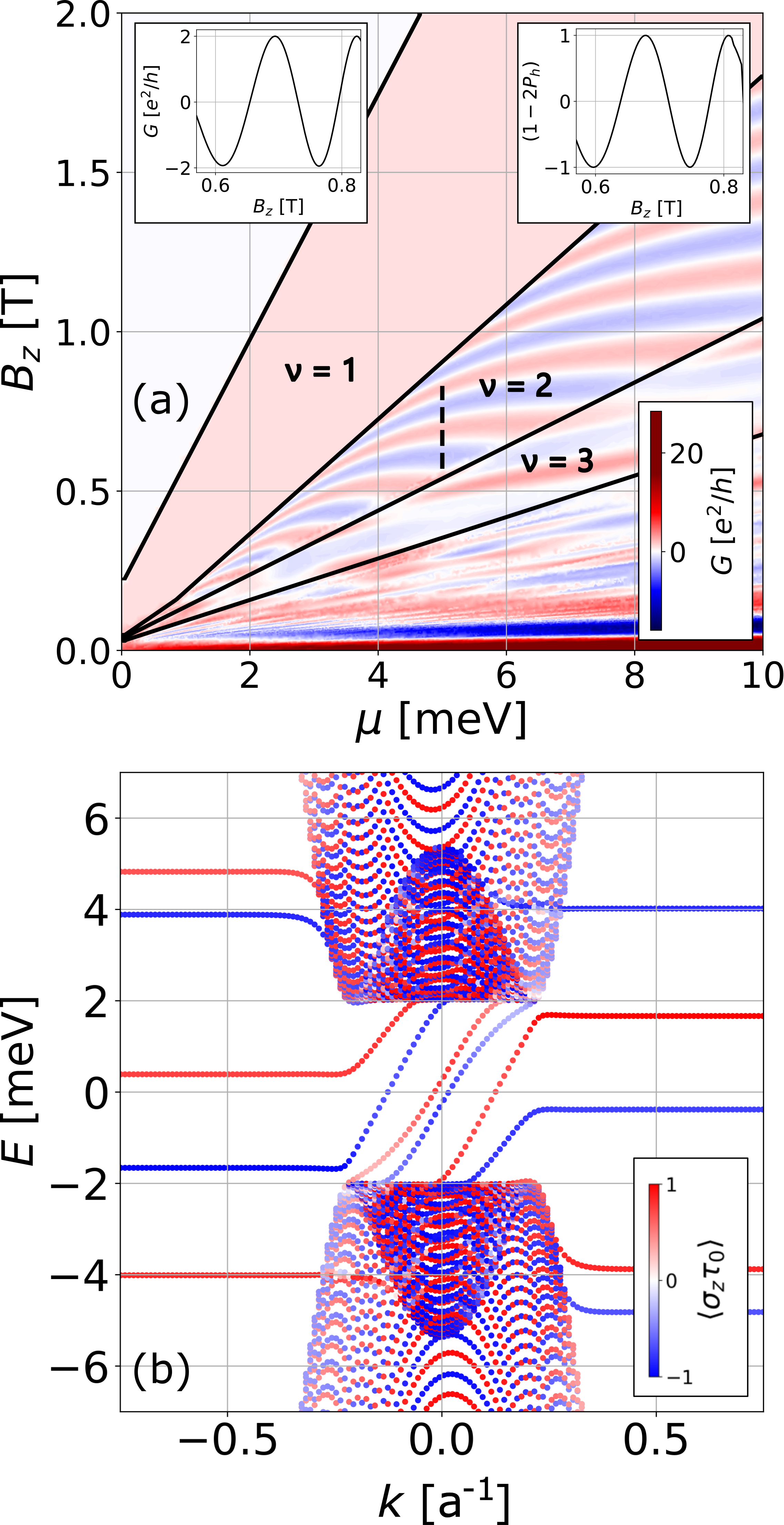}
    \caption{(a) Conductance map for varied perpendicular magnetic field and chemical potential with Zeeman interaction and spin-orbit coupling included  with $g=-50$, $\alpha = 50$ meVnm. The left inset of the map shows conductance cross-section versus $B_z$ at $\mu=5$ meV, right one shows the probability calculated from the analytical formula Eq.~\ref{eqn:probability}. (b) The band structure of the system colored by spin-polarization of the bands at $B_z = 0.6$ T, $\mu = 5$ meV. The results are obtained for $\alpha = 50$ meVnm, $g=-50$.}
    \label{fig:G_Zeeman_z_SO}
\end{figure}

\begin{figure}
    \centering
    \includegraphics[width=0.9\linewidth]{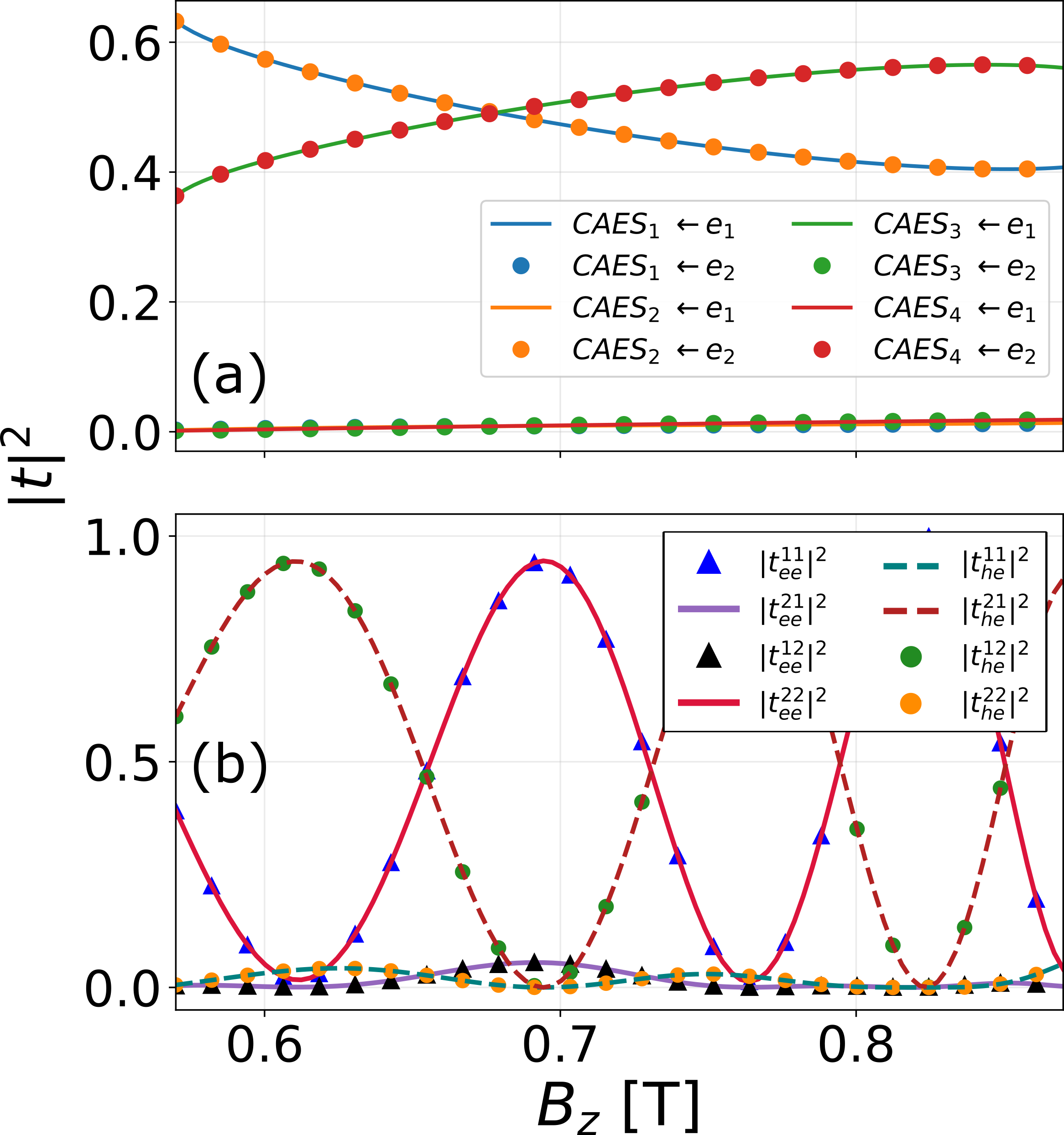}
    \caption{(a) Occupation probability of different CAES states for different initial electron edge modes in the $\nu = 2$ regime in the presence of spin-orbit coupling and perpendicular magnetic field with weak Zeeman interaction with $g=-50$, $\alpha = 50$ meVnm. (b) Components of electron and hole transmission probability. The results are obtained for perpendicular magnetic field with Zeeman and SOI interactions included with $g=-50$, $\alpha = 50$ meVnm and $\mu = 5$~meV.}
    \label{fig:transmission_probability_B_z_SO}
\end{figure}

Spin-orbit coupling alternates the spin of the charge carriers by coupling them to their momenta.  In Fig.~\ref{fig:G_Zeeman_z_SO}(a) we plot the conductance map of a QH-superconductor system with the inclusion of the Rashba SOI in the normal region. We observe that, in addition to a small shift in the chemical potential at which the regions of given $\nu$ appear, the conductance behavior is similar to that with the pure Zeeman interaction of Fig. \ref{fig:G_Zeeman_z}(a). The spin-mixing introduced by SOI causes a slight spin depolarization of CAES, which can be observed in the dispersion relation of Fig. \ref{fig:G_Zeeman_z_SO}(b). Nevertheless, the conductance oscillations in the $\nu = 2$ regime, where only the first edge mode is populated, still do not exhibit the mixing of CAES from the two spin sectors. This can be further inspected by studying the transmission probabilities from the normal lead edge modes of different spin polarizations ($e_1$ and $e_2$) in Fig. \ref{fig:transmission_probability_B_z_SO}. We observe that the electron from the band of a given spin polarization couples only to CAES of the same spin polarization. The probability of coupling to the opposite spin polarization is minimal, which protects the CAES from mixing, and the correspondence between the analytically extracted probabilities of electron conversion into hole and the numerically obtained conductance oscillations is preserved [cf. the insets to Fig. \ref{fig:G_Zeeman_z_SO}].

\begin{figure}
    \centering
    \includegraphics[width=0.85\linewidth]{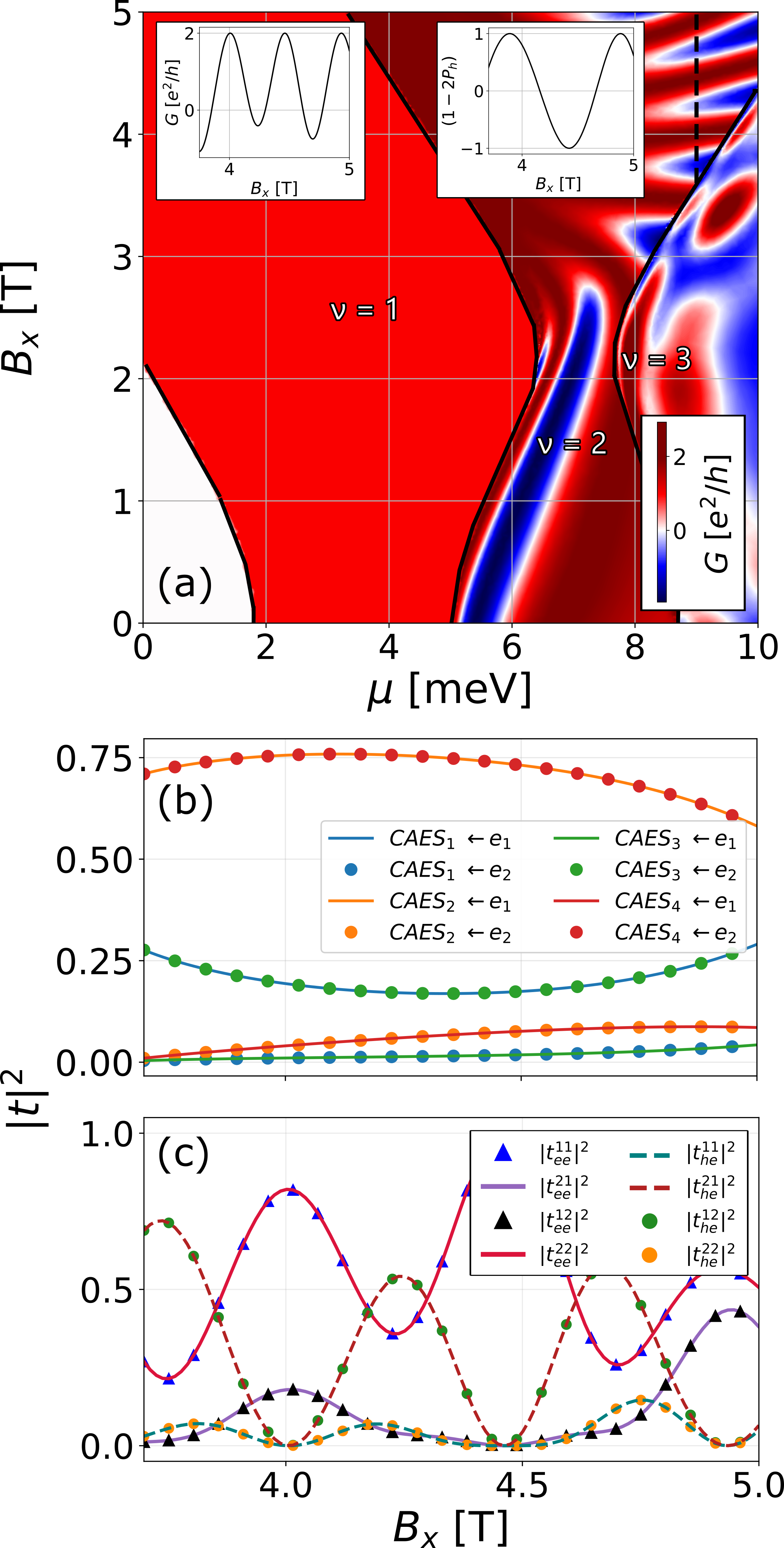}
    \caption{(a) Conductance of a QH-superconductor system versus chemical potential and the magnitude of in-plane magnetic field oriented in $x$-direction with spin-orbit coupling included with $g=-50$, $\alpha = 50$ meVnm. The left inset of the map shows conductance cross-section versus $B_x$ at $\mu=9$ meV, right one shows the probability calculated from the analytical formula Eq. \ref{eqn:probability}. (b) The transmission probabilities of an injected electron ending up in any one of the CAES at $\mu=9$ meV with $g=-50$, $\alpha = 50$ meVnm. (c) The corresponding probabilities of electron to electron and electron to hole transmission probabilities. The results are obtained for $B_z = 0.912$ T and $\alpha = 50$ meVnm,  $g=-50$.}
    \label{fig:G_Zeeman_x_SO}
\end{figure}

The situation for the in-plane magnetic field with included SOI is, however, significantly different. When we compare the conductance map obtained in the absence of SOI and the varied $B_x$ presented in the map of Fig. \ref{fig:G_Zeeman_x}(a) to the one obtained in the presence of SOI in Fig. \ref{fig:G_Zeeman_x_SO}(a), we observe that the $\nu = 2$ region does not close and reopen without conductance oscillations as $B_x$ is raised. The crossing of regions with different $\nu$ values is rather replaced by an hourglass shape with a constant $\nu = 2$ value inside. In fact, as $B_x$ is increased, the strong Zeeman interaction polarizing the spins in the $x$-direction splits the electron and hole bands in energy. Without SOI, when the splitting is strong enough, we are left with two electron (hole) bands with the same spin polarization below the Fermi energy; hence, no Andreev reflection can occur, which corresponds to the region of constant conductance marked with the green $\nu =2$ symbol in Fig. \ref{fig:G_Zeeman_x}(a). In the presence of SOI, however, the spin mixing does not allow for ideal polarization of the spins in the two remaining bands. This opens the possibility for Andreev reflection and hence CAES creation. As a result, we do not see constant conductance in the $\nu = 2$ regime at larger $B_x$ as observed previously. Here, in fact, below the Fermi energy, we have two orbital edge modes. As we learned previously, Andreev reflection introduces mixing between such modes, and hence now all four CAES mix. This can be inspected when looking at the transmission probability from the two available electron edge modes to the four CAES shown in Fig. \ref{fig:G_Zeeman_x_SO}(b). Clearly, there is considerable probability of coupling to all four CAES modes for every incoming electron mode, which is a signature of CAES mixing. As a signature of this mixing, we observe that in the large $B_x$ regimes, the analytical prediction assuming two uncoupled CAES bands does not reproduce the numerically obtained conductance oscillations [see the insets to Fig. \ref{fig:G_Zeeman_x_SO}(a)]. It is worth noting that for smaller values of the $B_x$ field, when the two electron bands correspond to different spin polarizations but the same edge mode number, the mixing is minimal in the same way as for the case of the pure $B_z$ field of Fig. \ref{fig:transmission_probability_B_z_SO}.

\begin{figure}
    \centering
    \includegraphics[width=0.85\linewidth]{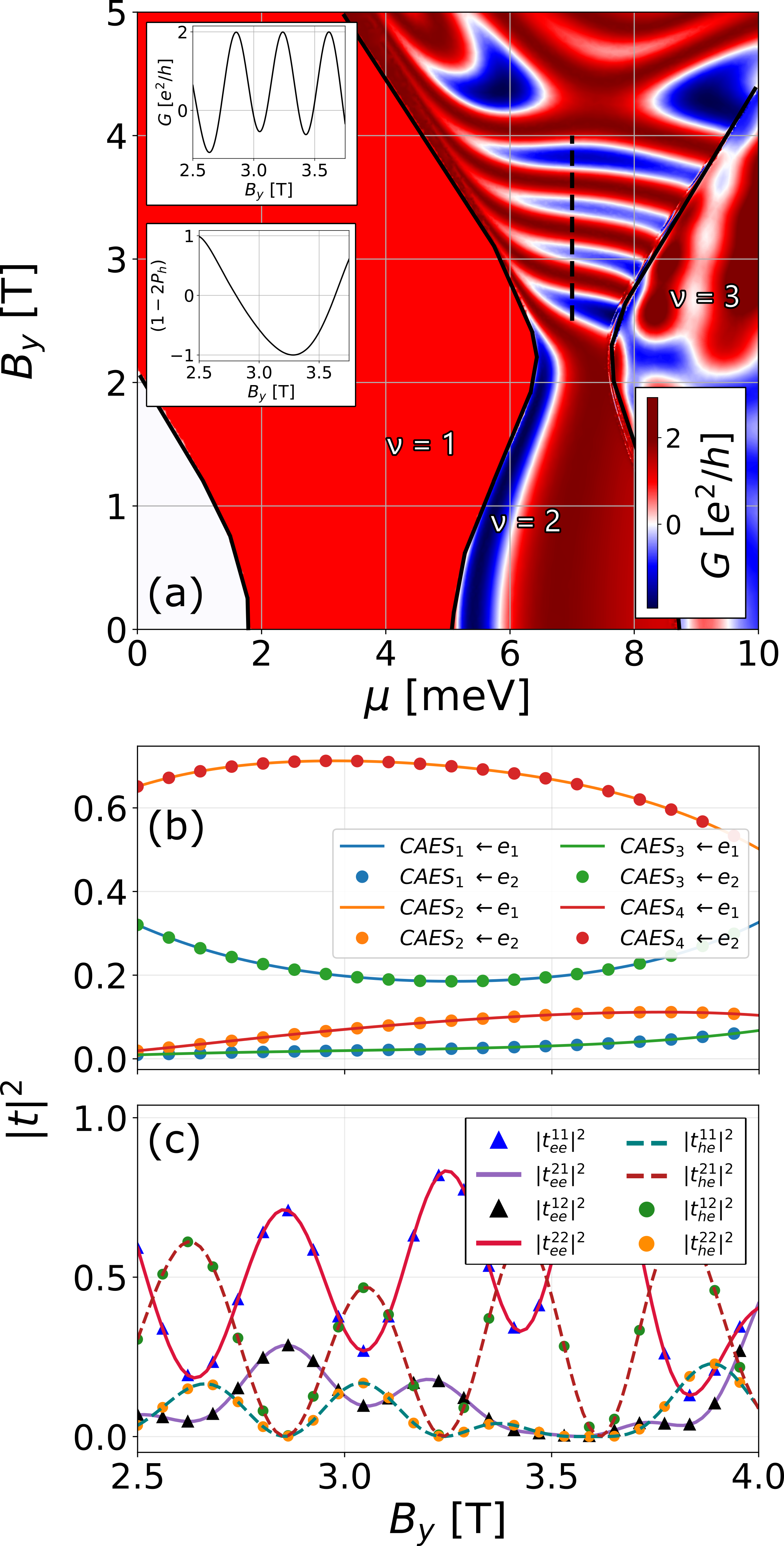}
    \caption{(a) Conductance of a QH-superconductor system versus chemical potential and the magnitude of in-plane magnetic field oriented in $y$-direction with spin-orbit coupling included with $g=-50$, $\alpha = 50$ meV nm. The left inset of the map shows conductance cross-section versus $B_y$ at $\mu=7$ meV, right one shows the probability calculated from the analytical formula \ref{eqn:probability}. (b) The transmission probabilities of an injected electron ending up in any one of the CAES at $\mu=7$ meV,  $g=-50$, $\alpha = 50$ meV nm. (c) The corresponding probabilities of electron to electron and electron to hole transmission probabilities. The results are obtained for $B_z = 0.912$ T and $\alpha = 50$ meVnm, $g=-50$. }
    \label{fig:G_Zeeman_y_SO}
\end{figure}

Similarly to the case of the in-plane magnetic field oriented in the $x$ direction, the in-plane field in the $y$-direction also introduces mode mixing. However, now that SOI is present, the conductance oscillations are sensitive to the orientation of the in-plane field, as the SOI interaction breaks the spin-rotational symmetry of our system. The presence of the field oriented in the $y$-direction produces a different pattern of oscillations that can be seen by comparing Fig. \ref{fig:G_Zeeman_x_SO}(a) to Fig. \ref{fig:G_Zeeman_y_SO}(a), as well as a different distribution of the transmission probabilities.

\subsubsection{Transmission degeneracies}

In Figs. \ref{fig:G_Zeeman_x_SO}(b) and \ref{fig:G_Zeeman_y_SO}(b), we observe striking symmetry. Every value of the transmission probability is degenerate; i.e., there are always two transport processes that occur with the same probability. For instance, we observe that the probability of an electron injected in the first mode and coupled to the second CAES mode (orange curve) is exactly the same as the probability of an electron injected in the second mode being coupled to the fourth CAES mode (red dots). Similar symmetries can be found for all other electron to hole transmission pairs. This is surprising because, at first glance, there is no similarity between the two electron or two CAES modes; they have different spatial profiles and spin polarizations due to the mixing provided by spin-orbit coupling. 

To understand this behavior, let us look at the full transmission matrix of the system (rounded to three significant figures) obtained for one exemplary set of parameters ($\mu = 9$ meV, $B_z = 0.912$ T, $B_x = 3.75$ T):
\begin{widetext}
\begin{equation}
    \mathcal{T}=\begin{pmatrix}
    0.392-0.318i & 0.071+0.022i & -0.24-0.816i & -0.08 + 0.097i  \\
    0.458+0.717i & -0.05 + 0.115i & -0.498+0.087i & 0.052 + 0.054i \\
    -0.052+0.054i & 0.498+0.087i & 0.05 + 0.115i & -0.458+0.717i \\
    0.08+0.097i & 0.24-0.816i & -0.071+0.022i & -0.392-0.318i
    \end{pmatrix}.
\end{equation}
\end{widetext}

The matrix is written in such a way that the columns number the incoming electron and hole modes, and the rows enumerate the CAES modes. Here we find all the degeneracies found in Fig. \ref{fig:G_Zeeman_x_SO}(b), e.g. $|\mathcal{T}_{2,1}|^2 = |\mathcal{T}_{4,2}|^2$. In the mode ordering assumed here, particle hole symmetry for the incoming modes connects the first (second) electron mode and the second (first) hole mode. The same symmetry holds for the CAES modes. At zero energy, this guaranties $\mathcal{T}_{i,j} = -\mathcal{T}^*_{{\bar{i},\bar{j}}}$, where $j$ denotes the incoming and $i$ the outgoing mode and $\bar{i}$ denotes the particle-hole partner of mode $i$. Similarly, the transport probabilities are equal when swapping the incoming and outgoing modes with their particle-hole partners. This, by itself, however, does not explain the symmetry obtained in Fig. \ref{fig:G_Zeeman_x_SO}(b).

Since we are dealing with a QH system, where backscattering is not possible, the transmission matrix contains all the information of the full scattering matrix and is unitary. The unitarity guaranties that the sum of every column or row of the $|\mathcal{T}|^2$ matrix is equal to one. Taking advantage of this property, we can compare the sum of the first row and first column of the transmission probabilities, and using PHS symmetry of the matrix, we arrive at the equality $P_{1,2} + P_{1,3} = P_{2,1} + P_{2,4}$ where $P_{ij}$ is defined in Eq. \ref{eqn:probability_CAES}. This equation shows that the sums of the probabilities balance, but it does not mandate that the individual components are equal to each other. To break this degeneracy, we must invoke the off-diagonal orthogonality constraint of the unitary matrix $\mathcal{T}$, which is encapsulated in Jacobi's complementary minor formula:
\begin{equation}
\det(\mathcal{T}_J^I) = (-1)^{\sum I + \sum J}\det \mathcal{T} \det \left( (\mathcal{T}^{-1})^{\bar{J}}_{\bar{I}}\right),
\end{equation}
where $\mathcal{T}_J^I$ and $(\mathcal{T}^{-1})^{\bar{J}}_{\bar{I}}$ are complementary minors of the original transmission matrix. $I$ and $J$ denote the rows and columns chosen from the original matrix to create the minor.

Using the unitarity condition of the $\mathcal{T}$ matrix, we can write
\begin{equation}
\mathcal{T}^{I=1,4}_{J=2,3} = 
\begin{pmatrix}
\mathcal{T}_{1,2} & \mathcal{T}_{1,3}\\
\mathcal{T}_{4,2} & \mathcal{T}_{4,3}
\end{pmatrix} = B
\end{equation}
and
\begin{equation}
(\mathcal{T}^{-1})^{\bar{I}}_{\bar{J}} = (\mathcal{T}^{\dagger})^{I = 1,4}_{J=2,3} = 
\begin{pmatrix}
\mathcal{T}_{2,1}^* & \mathcal{T}_{3,1}^*\\
\mathcal{T}_{2,4}^* & \mathcal{T}_{3,4}^*
\end{pmatrix} = C^*,
\end{equation}
arriving at equation $\det(B) = \det(\mathcal{T})\det(C^*)$. Calculating the determinants, we obtain $-P_{1,2} + P_{1,3} = \det(\mathcal{T})(P_{2,4}-P_{2,1})$.  For the scattering matrix of our system in the presence of a strong in-plane magnetic field and considerable SOI, its determinant is purely real and equal to -1. Combining the resulting relation between the transmission coefficients with the one obtained from the unitarity of the transmission matrix and solving the resulting system of linear equations, we obtain $P_{1,2} = P_{3,1}$ and $P_{2,1} = P_{4,2}$, which directly correspond to the symmetries between electron and CAES mode transport observed in Figs. \ref{fig:G_Zeeman_x_SO}(b) and \ref{fig:G_Zeeman_y_SO}(b).

In the spinless case of Fig. \ref{fig:QH_SC_nu_2}, we do not observe such degeneracies. There, the system is spin degenerate, and in the absence of spin interactions, its Hamiltonian has a block diagonal form. It is then enough to consider only one block (i.e., corresponding to the basis elements $\Psi_{e {\uparrow}}$, $\Psi_{h{\downarrow}}$). The scattering matrix for such a system lacks consistent $\mathcal{T}_{j,i} = -\mathcal{T}^*_{{\bar{j},\bar{i}}}$ symmetry, and following Jacobi's determinant theorem, we only recreate the PHS relation between the amplitudes, which does not protect the symmetries in the transmission coefficients, resulting in distinctly different probabilities for electron modes seen in Fig. \ref{fig:QH_SC_nu_2}(c). 

Subsequently, for the results of the system in the presence of a sole magnetic field in the $z$-direction and SOI, we also find a degeneracy of the transmission probabilities from the electron modes to CAES [see Fig. \ref{fig:transmission_probability_B_z_SO}(a)]. The symmetries here are, however, different. The probability of occupation of the first CAES mode for an electron in the first mode is equal to the probability of the occupation of the second mode when the electron is injected from the second mode. This is a direct consequence of the CAES mode reordering when the magnetic field is lowered, which reorders the rows of the matrix and consequently changes the sign of the determinant to $\det(\mathcal{T})  = 1$. The same derivation performed above, considering the determinant equal to 1, leads to the degeneracies found in Fig. \ref{fig:transmission_probability_B_z_SO}(a), e.g., $P_{3,1} = P_{4,2}$, and so on. 

Finally, from those symmetries found for spinful cases, we have naturally emerging symmetries in electron to electron and electron to hole transmission probabilities, as observed in Figs. \ref{fig:G_Zeeman_x_SO}(c) and \ref{fig:G_Zeeman_y_SO}(c). Each electron (hole) to electron (hole) transport process (from a mode $j$ to mode $i$) can be decomposed into the process of an electron transmitted to CAES and then from CAES to an electron, with corresponding transmission probabilities. We find that for the cases of $i = j$, or $i=1,j=2$ and $i=2,j=1$, the sum of the product of those probabilities is the same; hence the degeneracies found in Figs. \ref{fig:G_Zeeman_x_SO}(c) and \ref{fig:G_Zeeman_y_SO}(c).

\section{Summary and Conclusions}
We theoretically investigated the transport properties and mode-mixing of chiral Andreev edge states in a superconductor-proximitized two-dimensional quantum Hall system using the Bogoliubov-de Gennes equations. In the spin-degenerate case, $\nu=2$ conductance oscillations follow simple two-mode interference. However, at higher filling factors (e.g., $\nu=4$), Andreev reflection fundamentally mixes the incoming quantum Hall edge modes, yielding a non-zero probability for electrons to escape in different modes---a process strictly prohibited in clean electronic systems.

Introducing Zeeman splitting separates energy bands into orthogonal spin species. Because chiral Andreev states form from opposite-spin electron-hole pairs (consistent with s-wave pairing), this orthogonality prevents mixing and preserves the two-mode interference model, eventually suppressing Andreev edge states creation entirely at strong fields. Conversely, combining Rashba spin-orbit coupling with an in-plane magnetic field disrupts this ideal spin polarization. This persistent spin mixing re-enables Andreev reflection, driving complex hybridization among all four chiral Andreev edge states bands and fundamentally altering the conductance oscillations.

Finally, we observed robust degeneracies in the transmission probabilities. We showed that these symmetric transport features are not coincidental, but emerge as direct consequences of the transmission matrix unitarity and the symmetry of transmission matrix elements due to particle-hole symmetry at zero bias.

\section{Acknowledgments}
SM acknowledges helpful discussions with Alfredo Levy Yeyati. This work was supported by the National Science Center, Poland (NCN) agreement number UMO-2020/38/E/ST3/00418. We gratefully acknowledge the Polish high-performance computing infrastructure PLGrid (HPC Center: ACK Cyfronet AGH) for providing computer facilities and support within the computational grant no. PLG/2025/018486.

\bibliography{reference}
\end{document}